# Emergent electromagnetic induction beyond room temperature

Aki Kitaori[1]*, Naoya Kanazawa[1]*, Tomoyuki Yokouchi[2], Fumitaka Kagawa[1,3],

Naoto Nagaosa[1,3], Yoshinori Tokura[1,3,4]*

[1]Department of Applied Physics, The University of Tokyo; Tokyo, 113-8656, Japan.
[2]Deapartment of Basic Science, The University of Tokyo; Tokyo, 152-8902, Japan.
[3]RIKEN Center for Emergent Matter Science (CEMS); Wako, 351-0198, Japan.
[4]Tokyo College, The University of Tokyo; Tokyo, 113-8656, Japan.

*Corresponding author. Email: kitaori-aki055@g.ecc.u-tokyo.ac.jp, kanazawa@ap.t.u-tokyo.ac.jp, and tokura@riken.jp

**Abstract**
Emergent electromagnetic induction based on electrodynamics of noncollinear spin states may enable dramatic miniaturization of inductor elements widely used in electric circuits, yet many issues are to be solved toward application. One such problem is how to increase working temperature. We report the large emergent electromagnetic induction achieved around and above room temperature based on short-period ($\leq$ 3 nm) spin-spiral states of a metallic helimagnet YMn$_6$Sn$_6$. The observed inductance value $L$ and its sign are observed to vary to a large extent, depending not only on the spin helix structure controlled by temperature and magnetic field but also on the current density. The present finding on room-temperature operation and possible sign control of $L$ may provide a new step toward realizing microscale quantum inductors.

**Main Text**

Conventional inductor based on classical electromagnetism is one of the most important elements in electric circuits, as characterized by the relation $V = L\ dI/dt$, where $V$, $I$, and $L$ are voltage, current and inductance, respectively. Since $L$ of the inductor coil is proportional to the product of square of the coil's winding number and the coil's cross-section, it is difficult to reduce the dimensions of the inductor while keeping $L$ large enough. To overcome the size problem of the coil-shaped inductor, the simple scheme of the electromagnetic induction has recently been proposed to use the current-induced spin dynamics in a helical-spin system (*1*) and experimentally verified for the helimagnetic phases of a metallic compound (*2*). The idea is to utilize the time-dependent emergent electromagnetic field or dynamics of Berry phase (*3, 4*) produced by the conduction electrons flowing on the helical spin texture[1].

The emergent magnetic field (**b**) acting on conduction electrons is realized in a non-coplanar spin texture endowed with scalar spin chirality, typically in the skyrmion-lattice phase (*5, 6*). For example, the current-driven motion of skyrmions accompanying **b** produces the emergent electric field (**e**), which is known to give a current-dependent correction to the topological Hall effect in the skyrmion-lattice phase (*7*). The generalized Faraday's law (*8*) tells that $\nabla \times \mathbf{e} = -\partial \mathbf{b}/\partial t$ or $\mathbf{e} = -\partial \mathbf{a}/\partial t$, where **a** is the Berry connection satisfying that $\mathbf{b} = \nabla \times \mathbf{a}$. Thus, to generate the emergent electric field on the spin helix, the originally static **b** is not necessarily required but only the time dependent deformation of the spin helix by ac electric current is sufficient. In consideration of spin transfer torque on spin helix in the continuum limit, the coordinate component of emergent electric field ($e_i$) is described as (*4, 5, 8*)

$$e_i = \frac{h}{2\pi e}\mathbf{n} \cdot (\partial_i \mathbf{n} \times \partial_t \mathbf{n}), \tag{1}$$

where **n**, $h$ and $e$ are a unit vector parallel to the direction of spins, Planck's constant and bare electron charge, respectively. As opposed to **b**, **e** is related to the dynamics of spin structures and proportional to the solid angle dynamically swept by $\mathbf{n}(t)$. Hence, the motion of non-collinear spin structures can induce $e_i$ (*9-14*). For example, in the case of a proper-screw (Bloch-wall like) helix (Fig.1A), the emergent electric field can be described as (*1*):

$$e_x = \frac{Ph}{e\lambda}\partial_t \phi, \tag{2}$$

where $\lambda$ is the period of helix, $P$ is a spin polarization factor, $\phi$ is the tilting angle of the spin from the spiral plane, and the *x*-axis is taken parallel to the magnetic modulation vector (**q**). We note that $e_x$ can be generated regardless of the direction of the helical plane, e.g. also in cycloidal-type (Néel-wall like) spin modulations.

The appearance of the emergent electromagnetic inductance was recently confirmed for the helimagnet Gd$_3$Ru$_4$Al$_{12}$ (*2*), in which various non-collinear spin structures, such as proper-screw and transverse conical (see Fig.1D and E), show up below 20 K with a short helical pitch of λ ~ 2.8 nm due to the magnetic frustration effect of the localized Gd moments coupled via Ruderman-Kittel-Kasuya-Yosida (RKKY) interaction (*15-19*). Several important features of the emergent inductor were experimentally confirmed or clarified by this first experimental demonstration (*2*): (1) the inversed size scaling law, as anticipated, that *L* for the emergent inductor increases with the reduction of the element cross section *S*; (2) the inherently negative sign of the emergent inductance *L* in contrast to the positive value for conventional classic inductors; (3) the highly nonlinear behaviour of the emergent inductance *L* with the current density; and (4) the frequency dependence of the emergent inductance showing the Debye-type relaxation feature. Whether these features of the emergent electromagnetic inductance are generic and common to all the possible spin-helix states, in particular to the room-temperature helix states, is an important subject to be verified by experiments, in addition to the possible further enhancement of the inductance value at higher temperatures.

The main purpose of the present study is to realize the large enough magnitude, *e.g.* exceeding micro-henry level, of the emergent electromagnetic inductance in the micrometre-sized device around and above room temperature; this would be a primarily step toward the applications of the emergent electric field to the actual electronic devices. As the natural extension of the related study (*2*), we sought for the high-temperature helimagnetic state with nanometre-scale spin-helical pitch in the metallic compound. Among several candidate materials, we target here the helimagnet $YMn_6Sn_6$ which comprises magnetic Mn kagome-lattice, as shown in Fig. 1B, and undergoes the helimagnetic transition approximately below 330 K (Fig. 1H) (*20-24*). The compound shows the short-period helimagnetic states (*25-27*) including proper-screw helix (H, Fig. 1D) and transverse conical state (TC, Fig. 1E), whose magnetic modulation vectors **q** run parallel to the *c*-axis, *i.e.* normal to the Mn kagome-lattice plane. As increasing the magnetic field applied parallel to the *a*-axis, *i.e.* normal to the **q** direction or *c*-axis, the H state turns into TC, and (at low temperatures through the fan like state (FL, Fig. 1F)) finally to the forced ferromagnetic state (FF, Fig. 1G), as shown in the phase diagram in the temperature vs. magnetic-field plane (Fig. 1H). Here, the phase diagram was obtained by the magneto-transport measurements on the actual micrometer scale device (see the inset of Fig. 1H) with referring to the results reported for the bulk crystal and the magnetic-phase assignments done in previous reports (*25*); the phase transition temperatures and critical magnetic fields show a good agreement between the micrometre-sized device and the bulk crystal of $YMn_6Sn_6$; see Fig. S2 in Supplementary Materials (SM).

It is to be mentioned here about the complex helical magnetic structures, in particular for the H and TC phases, appearing in this compound at relatively low magnetic fields and relatively high temperatures. As partly elucidated by recent neutron scattering studies (*26-28*), there may be coexisting plural *q*-modulations in the H and TC phase. The *q*-values (corresponding to λ values less than 2.6 nm) appear to be plural, such as two- or three-fold; for example, there appear coexistent multiple magnetic Bragg satellites from incommensurate *q*-states, and the relative weight of the respective *q*-states appears to change depending on each phase, temperature, or magnetic field (*26-28*). Such a complex feature and variation of the spin helix modulation may influence the variation of the emergent inductance magnitude or even its sign, yet the elaborate arguments may have to await a future study to fully clarify the detailed spin structures. Nonetheless, the very short helical pitches (large *q* values) are obviously favourable for the generation of high inductance value, as argued in Eq. (2). Below, we examine the characteristics of the emergent electromagnetic inductance for the micrometre-scale devices made of $YMn_6Sn_6$ on which the electric current is applied along the *c*-axis (i.e. parallel to the **q** vector) while changing temperature and magnetic field applied along the *a*-axis (normal to the *c*-axis), as shown in the inset of Fig. 1H.

Figure 2A exemplifies the imaginary part of ac resistivity ($\rho^{1f}$) at 270 K measured using the ac input current density $j = j_0 \sin(2\pi f t)$ ($j_0 = 2.5 \times 10^4$ A/cm$^2$ and $f = 500$ Hz); the device (#1) size are 4.8 µm × 9.3 µm in cross section (*S*) and 25.0 µm in voltage-terminal distance (*d*). We also confirmed reproducibility; see Fig. S1 in SI. The real part of inductance value *L* can be directly related to Im$\rho^{1f}$ via the relation,

$$L = \mathrm{Im}\rho^{1f} d / 2\pi f S \qquad (3)$$

and plotted on the right ordinate scale for the present device. The Im$\rho^{1f}$ or *L* is mostly negative in the helimagnetic phases H and TC, in accord with the simplest theoretical prediction based on the spin-transfer torque mechanism when the collective spin excitation in spin helix state is gapped (>> *hf*, *h* being Planck's constant) (see the discussion in SM). The absolute value of *L* is nearly constant with magnetic field within the low-field H phase and increases in the TC phase, exceeding the value as large as 2 µH (micro-henry). In further increasing the magnetic field above 4 T, the absolute value of *L* steeply decreases within the TC phase, and around the TC to FF transition it changes the sign and shows a positive peak. As anticipated from the spin-helix origin of the emergent electric field generation, the *L* almost disappears when entering deeply the FF phase.

On the same device structure (#1) under the same current excitation condition ($j_0 = 2.5 \times 10^4$ A/cm$^2$), the temperature dependence of Im$\rho^{1f}(H)$ curve is shown in Figs. 2B-I together with the assignments of the magnetic phases (coloured vertical bands). The variation of Im$\rho^{1f}$ is also displayed in Fig. 2J as a colour contour map on the plane of temperature (*T*) and magnetic field (*H*). With lowering *T*, typically below 100 K, the absolute value of Im$\rho^{1f}$ is rapidly supressed, perhaps due to the increase of the *ab*-plane magnetic anisotropy which tends to suppress the current-induced spin deformation (amplitude in $\phi$ or $\partial_t \phi$ term in Eq. (2)). In the temperature region below 270 K, the Im$\rho^{1f}$ takes mostly negative values in the H, TC, and FL phases, while the magnitude is different in the respective phases and also depends on temperature. Notably, around the boundary between helimagnetic TC (or FL) and FF phases, the Im$\rho^{1f}$ once increases to take the positive-value peak (see the red-coloured region in Fig. 2J). The positive value of Im$\rho^{1f}$ or *L* is more clearly and broadly observed at higher temperatures above 300 K. For example, at 300 K the Im$\rho^{1f}$ is still negative within the H phase but shows a positive value in the whole TC phase, accompanying a sharp negative dip upon the field-induced H-to-TC transition. Surprisingly, even above the magnetic transition temperature (330 K), e.g. at 350 K, the broad positive peak is observed in low-field region, followed by the negative background in higher-field (> 3 T) region. One possible scenario to explain the positive sign of emergent inductance is the gapless feature of the spin collective mode. (A phenomenological model for the sign change of *L* is discussed in SM.) The sign change of *L* as observed around the phase boundary region at relatively high temperatures above 250 K (see Fig. 2J) suggests that the collective phason like mode originally associated with the energy gap due to the impurity pinning becomes gapless possibly due to the thermally induced depinning.

Next, we proceed to the frequency dependence and the current nonlinearity for the present emergent inductor. To investigate a wide range of frequency dependence, we performed the LCR-meter measurement on the two-terminal device (#2 with 5.5 μm × 1.8 μm in $S$ and 28.8 μm in $d$, see Fig. S1 in SM) at zero field. The inset to Fig. 3 exemplifies the frequency ($f$) dependence of the real and imaginary parts of the inductance $L$ at 100 K. A prototypical Debye-type relaxation behaviour is observed there with the characteristic frequency ($f_0$) around 1 kHz, at which Im$L$ shows a peak (*2, 28*) (see also Methods for the definition of Im $L$). The similar relaxation-type $f$ dependence of Re$L$ is observed at various temperatures, as shown in the main panel of Fig. 3. The characteristic frequency $f_0 \sim$ 1 kHz is rather insensitive to the temperature variation, while the $L$ value changes from negative to positive in approaching the magnetic transition temperature ($T_N \sim$330 K). This means that the deformation of spins cannot fully follow alternating currents with a frequency above 1 kHz, which is ascribed to the extrinsic pinning effect stemming from defects/impurities. As compared with the case of the low-$T_N$ (~ 20K) helimanget $Gd_3Ru_4Al_{12}$ where $f_0 \sim$ 10 kHz and λ ~ 2.8 nm (*2*), one order of magnitude lower $f_0$ in the present compound with comparable λ may be ascribed to the smaller extrinsic pinning effect. Such frequency dependence of the $L$ would be improved to show higher $f_0$ by enhancing the extrinsic pinning effect via, for example, artificially introducing the pinning sites.

As for the current-nonlinear behaviour of $L$, this compound shows dramatic but complex magnetic-field dependent features. When the current density $j$ is varied, the magnetic-field variation of the inductance or Im$\rho^{1f}$ is observed to change in a qualitative manner, even including its sign. Shown in Fig.4A is the typical result at 270 K on the device #3 (3.8 μm × 8.5 μm in $S$ and 35.5 μm in $d$, see Fig. S1 in SM); the magnetic-field dependence at the current density $j_0 \sim 2 \times 10^4$ A/cm$^2$ therein nearly reproduces the results of the device #1 shown in Fig. 2A at $j_0 = 2.5 \times 10^4$ A/cm$^2$. The H phase around zero field shows the negative Im$\rho^{1f}$, whereas within the TC phase between 2.2 T and 5.2 T the Im$\rho^{1f}$ shows a competitive behaviour between the originally negative and the $j$-accelerated positive components. To see this more clearly, we plot the evolution of Im$\rho^{1f}$ with the current density as monitored at 0 T (H phase), 2.5 T (lower-field side of TC phase) and 4.2 T (higher-field side of TC phase) in Fig. 4B together with the negative-maximal and positive-maximal magnitudes in this magnetic field range eye-guided by the envelope curves. At 0 T in the H phase, the Im$\rho^{1f}$ shows a clear nonlinear behaviour, but remains negative in the present current-density range. The measurement of the third harmonic signal Im$\rho^{3f}$ also confirmed the $j$-nonlinear behaviour; the third-order nonlinearity ($j^3$-term) for the spin-spiral plane distortion $\phi$ shows up as the coefficient of $j^2$-term in Im$\rho^{1f}$ as well as the Im$\rho^{3f}$ itself (*2*), which is observed to become dominant over the linear response already at $j_0 \sim 2 \times 10^4$ A/cm$^2$. In approaching the TC-FF boundary, *e.g.* at 4.2 T, the $j$-nonlinear change to the positive value becomes conspicuous, indicating that the expansion of Im$\rho^{1f}$ with higher-order polynomials of $j$ is no more valid. Instead, the current induced change of the magnetic structure itself, such as the weight change of the plural $q$-value components even within the TC phase region, should be taken into account for the highly $j$-nonlinear. This is to be confirmed by the in-situ neutron or magnetic resonant x-ray scattering studies while changing the magnetic field and electric current density. In turn, the possible control of the emergent inductance sign and magnitude in terms of the current excitation would be an important function for this class of quantum inductor.

We have demonstrated the potential of the above-room-temperature quantum inductor as derived from the emergent electric field generated in the short-pitch (~ 3 nm) helimagnetic states, such as proper-screw helix and transverse conical states, of the YMn$_6$Sn$_6$ crystal plate. The imaginary part of ac resistivity Im $\rho^{1f}$ as the representation of the intrinsic material parameter for the electromagnetic induction can show large absolute values of ~µΩcm level, ensuring the large inductance value for the µm-sized or thin-film device, in which the current density can be increased. The emergent inductance is highly nonlinear with the current density exceeding $10^4$ A/cm$^2$, mostly negative but turns into a large positive value near the phase boundary to the forced ferromagnetic region or with increasing the current density in the transverse conical spin state. The positive sign of the emergent inductance is likely due to the gapless nature of collective spin excitation, however the detailed mechanism for the magnetic-field induced sign change remains to be clarified. The possibility of the varying magnitude and sign of the emergent inductance with current density may lead to a useful functionality of the quantum inductor as well as the advantage of several orders-of-magnitude miniaturisation as compared with the classical coil inductor.

**References and Notes**

1. N. Nagaosa, Emergent inductor by spiral magnets. *Jpn. J. Appl. Phys.* **58,** 12090 (2019).
2. T. Yokouchi *et al*. Emergent electromagnetic induction in a helical-spin magnet. *Nature* **586**, 232-236 (2020).
3. M. V. Berry, Quantal phase factors accompanying adiabatic changes. *Proc. R. Soc. Lond. A* **392,** 45-47 (1984).
4. D. Xiao, M.-C. Chang, Q. Niu. Berry phase effects on electronic properties. *Rev. of Mod. Phys.* **82**, 1959-2006 (2010).
5. N. Nagaosa, Y. Tokura, Topological properties and dynamics of magnetic skyrmions. *Nat. Nanotechnol*. **8**, 899-911 (2013).
6. A. Neubauer *et al*. Topological Hall Effect in the A Phase of MnSi. *Phys. Rev. Lett.* **102**, 186602 (2009).
7. T. Schulz *et al.* Emergent electrodynamics of skyrmions in a chiral magnet. *Nature Phys.* **8,** 301-304 (2012).
8. G. E. Volovik, Linear momentum in ferromagnets. *J. Phys. C* **20**, L83-L87 (1987).
9. S. E. Barnes, S. Maekawa, Generalization of Faraday's Law to Include Nonconservative Spin Forces *Phys. Rev. Lett.* **98,** 246601 (2007).
10. S. A. Yang *et al.* Universal Electromotive Force Induced by Domain Wall Motion. *Phys. Rev. Lett.* **102,** 067201 (2009).
11. M. Hayashi *et al*. Time-Domain Observation of the Spinmotive Force in Permalloy Nanowires. *Phys. Rev. Lett.* **108,** 147202 (2012).
12. Y. Yamane *et al.* Continuous Generation of Spinmotive Force in a Patterned Ferromagnetic Film. *Phys. Rev. Lett.* **107**, 236602 (2011).
13. J. Kishine, I. G. Bostrem, A. S. Ovchinnikov, V. E. Sinitsyn, Coherent sliding dynamics and spin motive force driven by crossed magnetic fields in a chiral helimagnet. *Phys. Rev. B* **86,** 214426 (2012).
14. Y. Yamane, J. Ieda, J. Sinova, Electric voltage generation by antiferromagnetic dynamics. *Phys. Rev. B* **93**, 180408 (2016).
15. R. E. Gladyshevskii, O. R. Strusievicz, K. Cenzual, E. Parthé, Structure of $Gd_3Ru_4Al_{12}$, a new member of the $EuMg_{5.2}$ structure family with minority atom clusters. *Acta Crystallogr. B* **49,** 474-478 (1993).
16. J. Niermann, W. Jeitschko, Ternary rare earth (R) transition metal aluminides $R_3T_4Al_{12}$ ($T$ = Ru and Os) with $Gd_3Ru_4Al_{12}$ type structure. *J. Inorg. Gen. Chem.* **628,** 2549-2556 (2002).
17. S. Nakamura, *et al*. Spin trimer formation in the metallic compound $Gd_3Ru_4Al_{12}$ with a distorted kagome lattice structure. *Phys. Rev. B* **98,** 054410 (2018).
18. V. Chandragiri, K. K. Iyer, E. V. Sampathkumaran, Magnetic behavior of $Gd_3Ru_4Al_{12}$, a layered compound with distorted kagomé net. *J. Phys.: Condens. Matter* **28,** 286002 (2016).
19. M. Hirschberger *et al*. Skyrmion phase and competing magnetic orders on a breathing kagomé lattice. *Nat. Commun.* **10,** 5831 (2019).

20. G. Venturini, D. Fruchart, B. Malaman, Incommensurate magnetic structures of $RMn_6Sn_6$ (R = Sc, Y, Lu) compounds from neutron diffraction study. *J. Alloys Compd.* **236**, 102-110 (1996).

21. A. Matsuo, *et al*. Study of the Mn‐Mn exchange interactions in single crystals of $RMn_6Sn_6$ compounds with R= Sc, Y and Lu. *J. Alloys Compd.* **408‐412**, 110-113 (2006).

22. K. Uhlířová, *et al*. Magnetic properties and Hall effect of single-crystalline $YMn_6Sn_6$. J. Magn. Magn. Mater **310**, 1747-1749 (2007)

23. A. A. Bykov, Y. O. Chetverikov, A. N. Pirogov, S. V. Grigor'ev, Quasi-Two-Dimensional Character of the Magnetic Order‐Disorder Transition in $YMn_6Sn_6$. *JETP Lett.* **101** 699-702 (2015)

24. Q. Wang, Q. Yin, S. Fujitsu, H. Hosono, H. Lei, Near-room-temperature giant topological Hall effect in antiferromagnetic kagome metal $YMn_6Sn_6$. arXiv:1906.07986 (2019)

25. N. J. Ghimire *et al*. Competing magnetic phases and fluctuation-driven scalar spin chirality in the kagome metal $YMn_6Sn_6$. *Sci. Adv.* **6** 51 eabe2680 (2015)

26. K. J. Neubauer *et al*. In-plane magnetic field induced double fan spin structure with *c*-axis component in metallic kagome antiferromagnet $YMn_6Sn_6$. *Phys. Rev. B* **103**, 014416 (2021)

27. H. Zhang *et al*. Topological magnon bands in a room-temperature kagome magnet. *Phys. Rev. B* **101,** 100405 (R) (2020).

28. W. Kleemann, Universal domain wall dynamics in disordered ferroic materials. *Annu. Rev. Mater. Res.* **37,** 415-448 (2007).

29. K. Momma, F. Izumi, VESTA 3 for three-dimensional visualization of crystal, volumetric and morphology data. *J. Appl. Crystallogr.* **44**, 1272-1276 (2011).

**Acknowledgments**
The authors thank M. Kawasaki, Y. Kaneko and Y. Onishi for enlightening discussions. This work was supported by Core Research for Evolutional Science and Technology (CREST), Japan Science and Technology Agency (JST) (Grant No. JPMJCR1874 and No. JPMJCR16F1) and the Japan Society for the Promotion of Science (JSPS) KAKENHI (Grant No. JP20H01859 and No. JP20H05155).

**Funding**

Japan Science and Technology Agency (JST) CREST No. JPMJCR1874

Japan Science and Technology Agency (JST) CREST No. JPMJCR16F1

Japan Society for the Promotion of Science (JSPS) KAKENHI No. JP20H01859

Japan Society for the Promotion of Science (JSPS) KAKENHI No. JP20H05155

**Author contribution**

Conceptualization: YT, NN

Methodology: AK, NK, FK, TY, NN, YT

Investigation: AK, FK

Funding acquisition: YT, NN, NK

Project administration: YT, NN

Supervision: YT, NK

Writing – original draft: YT, NN

Writing – review & editing: AK, NK, FK, TY, NN, YT

**Competing interests**

Authors declare that they have no competing interests.

**Data and materials availability**

All data are available in the main text or the supplementary materials.

**Supplementary Materials**

Materials and Methods

Supplementary Text

Figs. S1 to S2

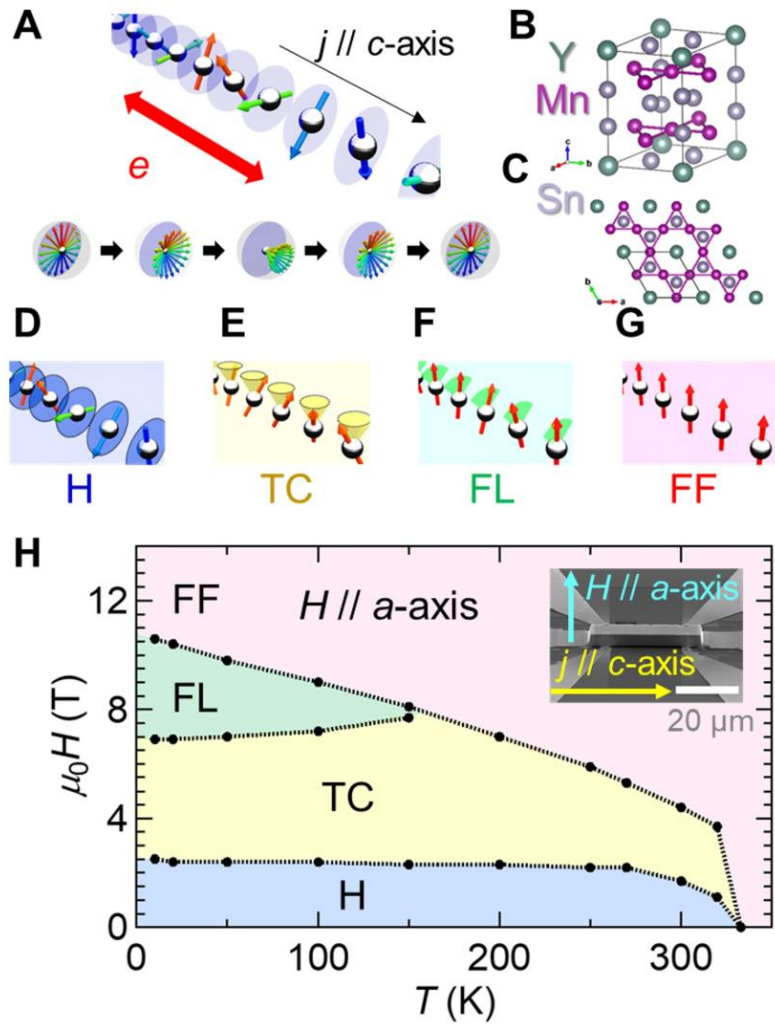

**Fig. 1. Conceptual diagram of emergent induction in spin helix and helimagnetic orders in YMn$_6$Sn$_6$. A**, Schematic illustration of emergent electromagnetic induction produced in the spin-helix state. Spin configurations of a proper-screw helix in real-space (top) and its projection into the unit sphere under the current-driven motion at several elapsed-time points (bottom). **B, C**, Oblique (B) and top (C) views of crystal structure of YMn$_6$Sn$_6$, built up from Y layers with triangular lattices, Mn layers with Kagome lattices and Sn layers with honeycomb lattices. **D-G**, Schematic illustrations of proper-screw helical (D), transverse conical (E), fan-like (F) and forced ferromagnetic (G) structures. H, YMn$_6$Sn$_6$ magnetic phase diagram for $H \parallel a$-axis. The blue, yellow, green, and red regions represent proper-screw helical (H), transverse conical (TC), fan-like (FL), and forced ferromagnetic (FF) phases, respectively. The inset is a scanning electron microscope (SEM) image of a YMn$_6$Sn$_6$ thin plate device.

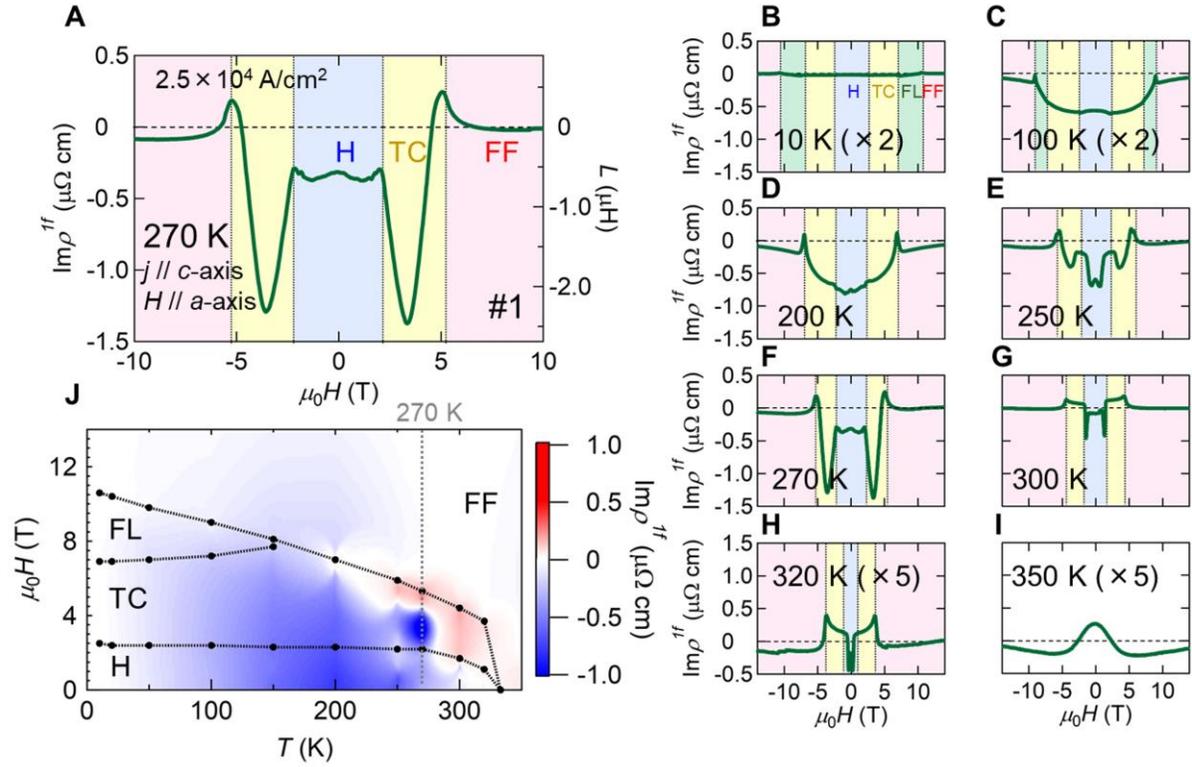

**Fig. 2. Emergent inductance beyond room temperature and its temperature and magnetic-field dependence in YMn$_6$Sn$_6$. A-I,** Magnetic-field ($H$) dependence of the imaginary part of the complex resistance Im$\rho^{1f}$ measured under $H \parallel a$-axis and an ac input current density $j = j_0 \sin(2\pi ft)$ ($j_0 = 2.5 \times 10^4$ A/cm$^2$, $f = 500$ Hz, $j \parallel c$-axis) at various temperatures. The blue, yellow, green and red shadows represent proper-screw helical (H), transverse conical (TC), fan-like (FL) and forced ferromagnetic (FF) phases, respectively. **J,** Colour contour mapping of Im$\rho^{1f}$ in the $T$-$H$ plane. Phase boundaries are indicated by black dots and dashed lines.

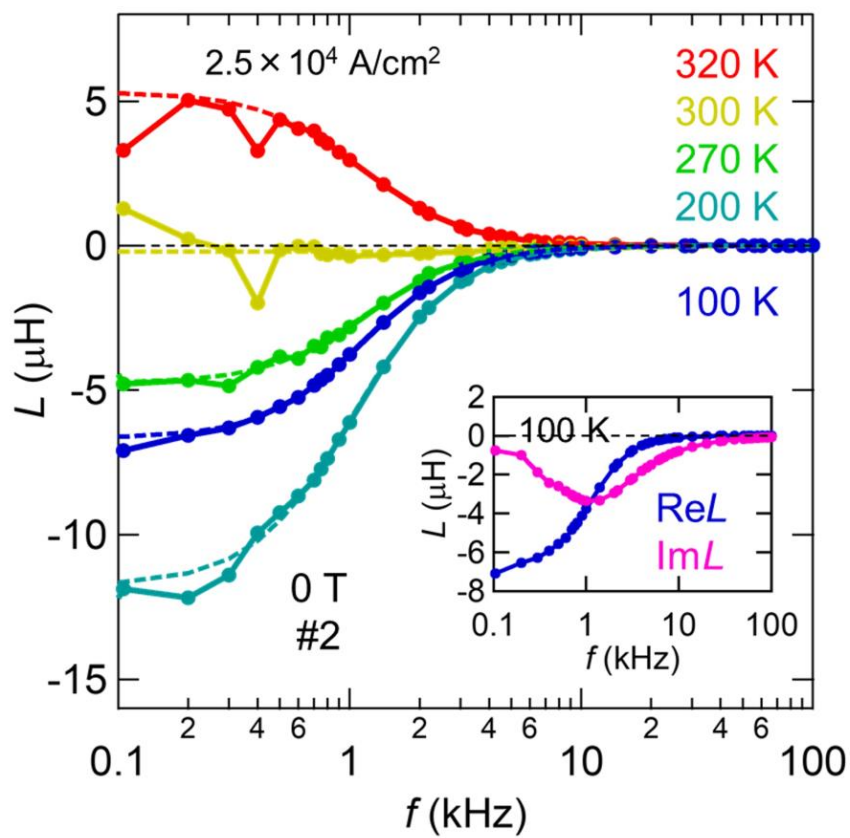

**Fig. 3. Frequency dependence of emergent inductance.** Frequency dependence of the real part of complex inductance measured with the LCR meter at zero magnetic field. The dashed lines are fitting curves assuming Debye-type relaxation. Inset shows a prototypical Debye-type relaxation behavior of complex inductance observed at 100 K.

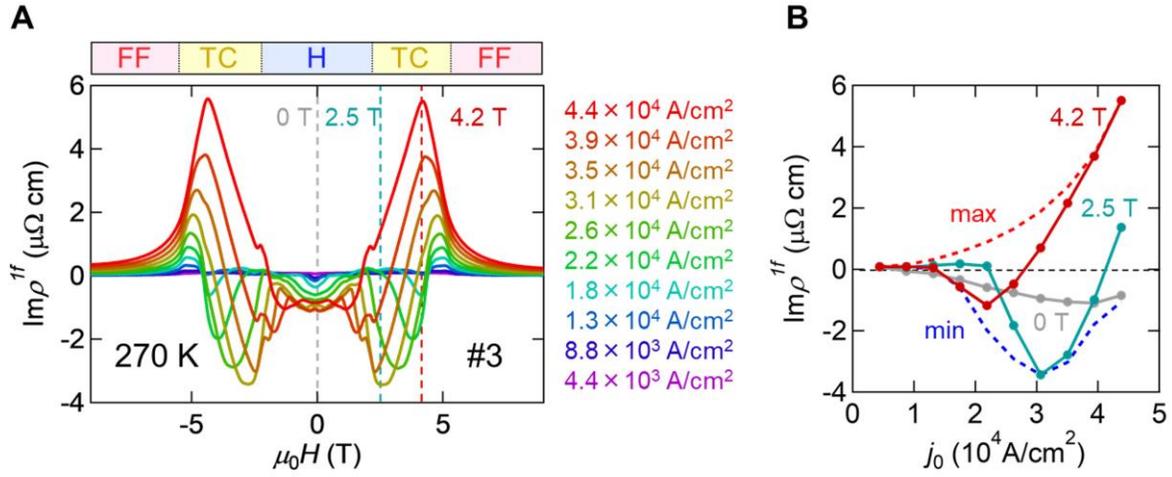

**Fig. 4. Nonlinear emergent inductance with several current densities at 270 K. A**, Magnetic field ($H$) dependence of the imaginary part of complex resistance $\mathrm{Im}\rho^{1f}$ measured under $H \,/\!/\, a$-axis and different amplitudes of a.c. current density $j = j_0 \sin(2\pi ft)$ ($f$ = 500 Hz, $j \,\|\, c$-axis). **B**, Current density ($j_0$) dependence of $\mathrm{Im}\rho^{1f}$ at specific $H$ values. Red and blue dashed lines represent positive- and negative-maximal magnitudes of $\mathrm{Im}\rho^{1f}$ at each $j_0$, respectively.

# Supplementary Materials for

Emergent electromagnetic induction beyond room temperature

Aki Kitaori[1]*, Naoya Kanazawa[1]*, Tomoyuki Yokouchi[2], Fumitaka Kagawa[1,3],

Naoto Nagaosa[1,3], Yoshinori Tokura[1,3,4]*

[1]Department of Applied Physics, The University of Tokyo；Tokyo, 113-8656, Japan.
[2]Deapartment of Basic Science, The University of Tokyo; Tokyo, 152-8902, Japan.
[3]RIKEN Center for Emergent Matter Science (CEMS); Wako, 351-0198, Japan.
[4]Tokyo College, The University of Tokyo; Tokyo, 113-8656, Japan.

*Corresponding authors. Email: kitaori-aki055@g.ecc.u-tokyo.ac.jp, kanazawa@ap.t.u-tokyo.ac.jp, and tokura@riken.jp

**This PDF file includes:**

> Materials and Methods
> Supplementary Text
> Figs. S1 to S2

**Materials and Methods**

Crystal growth and device fabrication.

The single crystals of $YMn_6Sn_6$ were synthesized by a Sn-flux method (*22*). A mixture of ingredient elements with atomic ratio of Y:Mn:Sn = 1:6:30 was put in an evacuated quartz tube and heated to 1050 °C, subsequently cooled slowly to 600 °C and then quenched to room temperature. Any remaining flux was centrifuged, followed by soaking in hydrochloric acid solution. The single crystallinity was indicated by the well-developed facet structures and was also confirmed by Laue X-ray diffraction. No impurity phase of the single crystal was detected by powder X-ray diffraction. We cut thin plates out of the single crystals by using the focused ion beam (FIB) technique (NB-5000, Hitachi). The thin plates were mounted on silicon substrates with patterned electrodes. We fixed the thin plates to the substrates and electrically connected them to the electrodes by using FIB-assisted tungsten-deposition. We made Au/Ti-bilayer electrode patterns by an electron-beam deposition method.

Transport and magnetization measurements.

Magnetic-field dependence of complex resistivity was measured with use of lock-in amplifiers (SR-830, Stanford Research Systems). We input a sine-wave current and recorded both in-phase (Re $V^{1f}$) and out-of-phase (Im $V^{1f}$) voltage with a standard four-terminal configuration. Background signals were estimated by measuring a short circuit where the terminal pads were connected by Au/Ti-bilayer electrode patterns. We subtracted the background signals from the measurement data. The possible temperature increase $\Delta T$ upon current excitation was checked by monitoring the temperature-dependent resistance value of the sample by passing the dc current; note that the average joule heating by the dc current density $j_{dc}$ is twice as large as that by the ac current density ($j_0$) of the same amplitude. In the current excitation corresponding to the case of $j_0$ ~ $3.6 \times 10^4$ A/cm$^2$, close to the maximal value used to obtain the result of Fig.4, the estimated temperature increase is $\Delta T = +2.5$ K at the base temperature of 270 K, indicating little influence of the heating on the current-induced effects discussed in this work.

Frequency dependence of complex inductance was measured with use of LCR meter (Agilent Technologies, E4980A). We employed two-terminal method to reduce parasitic impedance (the device #2 shown in Fig. S1). We corrected the contributions from the cables and the electrodes with a standard open/short correction procedure. We also subtracted the contributions from electrical contacts between the sample and the electrodes, which were estimated by measurements with low current density ($j_0 = 1.0 \times 10^3 \text{A/cm}^2$). Here, the observed complex impedance $\tilde{Z}(\omega)$ is the sum of frequency-independent resistance ($R$) and frequency-dependent reactance of complex inductance [$\omega\tilde{L}(\omega) = \omega\text{Re}L(\omega) + i\omega\text{Im}L(\omega)$]. The real and imaginary components of inductance can be estimated as $\text{Re}L(\omega) = \text{Im}\tilde{Z}(\omega)/\omega, \text{Im}L(\omega) = (\text{Re}\tilde{Z}(\omega) - R)/\omega$.

**Supplementary Text**

Devices used for emergent inductance measurements.

We fabricated and measured several devices with different shapes and electrode configurations to confirm reproducibility. Fig. S1 shows a list of part of the fabricated devices and the representative data of imaginary part of ac resistivity $\text{Im}\rho^{1f}$ at 270 K. The label numbers in Fig. S1 correspond to the device numbers described in the figures of main text (#1-#3) and Fig. S2 (#4). Here, the current density and frequency for measurements were $2.5 \times 10^4$ A/cm$^2$ and 500 Hz, respectively. Irrespective of large variation of sample shape, $\text{Im}\rho^{1f}$ exhibits similar magnetic-field profiles, although the difference in magnitude are discerned especially around the phase boundaries. Note that the device #2 is the two-terminal device for the LCR measurement.

Comparison of magnetic and transport properties between a bulk crystal and a thin-plate device.

In Fig. S2, we show resistivity properties of a bulk single crystal and a thin-plate device (#4, see Fig. S1D). Both the temperature and magnetic-field profiles of resitivity in the thin-plate device almost perfectly trace those in the bulk crystal. This validates little influence of the micro-device fabrication procedure on the electronic/magnetic states of YMn$_6$Sn$_6$, such as damage by exposure to Ga-ion beam during the FIB process and strain from the silicon substrate. The kink structures in magnetoresistivity (upper panels of Fig. 2B) coincide with the magnetic transitions inferred from the magnetization curves (lower panels of Fig. S2B); magnetization measurement was performed in Quantum Design PPMS-14 T with ACMS option. Thus, the magnetic phase diagram would be least affected by the microfabrication procedure, and was derived, as shown in Fig. 1H, with referring to the phase assignments reported in previsous studies (*25*).

Phenomenological model to describe the sign change of emergent inductance.

The inductance $L(\omega)$ derived from the imaginary part of the impedance $Z(\omega)$ as $L(\omega) = \text{Im} \, Z(\omega)/\omega$; $Z(\omega)$ is the inverse of the conductance $\sigma(\omega)$. Here we present a phenomenological model for the conductance due to the excitation of some low-lying collective modes. Let us assume that the displacement $x$ obeys the equation of motion

$$m(\ddot{x} + \gamma\dot{x}) + \frac{\partial V(x)}{\partial x} = F(t) \qquad (S1)$$

where $\dot{}$ means the time-derivative, $V(x)$ is the potential function for $x$, and $F(t)$ is the external driving force, e.g., electric field.

The linear response can be formulated by expanding $V(x)$ around its stable minimum, i.e. $x = 0$, and Eq.(S1) is reduced to

$$m(\ddot{x} + \gamma\dot{x} + \omega_0^2 x) = F(t) \qquad (S2)$$

The complex conductance $\sigma(\omega)$ is defined as $\dot{x} = \sigma(\omega)F$ by putting $F(t) = Fe^{i\omega t}$ as

$$\sigma(\omega) = \frac{i\omega}{m} \frac{1}{\omega_0^2 - \omega^2 + i\gamma\omega} \qquad (S3)$$

This expression describes the contribution from the low-lying collective mode with $\omega_0$ corresponding to its pinning frequency.

Since the system of our interest is metallic, the usual Drude term $\sigma_D$ should be added to Eq.(S3). In the present case, $|\sigma_D| \gg |\sigma(\omega)|$, and hence

$$Z(\omega) = (\sigma_D + \sigma(\omega))^{-1} \approx \sigma_D^{-1} - \sigma_D^{-2} \sigma(\omega) \qquad (S4)$$

Note here that the characteristic frequency of the Drude term is much higher than those relevant to the collective modes, i.e., the inverse of the microscopic relaxation time of the order of pico second, and hence $\sigma_D$ is regarded as a real constant.
Therefore,

$$L(\omega) = \mathrm{Im}\frac{Z(\omega)}{\omega} = -\sigma_D^{-2}\frac{\mathrm{Im}\sigma(\omega)}{\omega} = -\frac{\sigma_D^{-2}}{m}\frac{(\omega_0^2-\omega^2)}{(\omega_0^2-\omega^2)^2+(\gamma\omega)^2} \qquad (S5)$$

In Eq.(S5), it is seen that the inductance $L(\omega)$ is negative for $\omega < \omega_0$, while positive for $\omega > \omega_0$. This leads to the conclusion that we need the gapless collective excitation for the positive inductance in such a low frequency limit as the ac frequency in the experiment.

One needs to solve the equation of motion in eq.(S1) for the nonlinear response to $F$ in the generic choices of $V(x)$, However, one can give a generic discussion in the limit $\omega \to 0$ as follows. In this limit, one can drop the inertia term in Eq.(S1), and the equation becomes

$$m\gamma\dot{x} = -\frac{\partial V(x)}{\partial x} + F(t), \qquad (S6)$$

assuming the slowly varying $F(t)$. One can construct the solution to Eq. (S6) starting from the static solution, i.e.,

$$-\frac{\partial V(x)}{\partial x} + F = 0, \qquad (S7)$$

Let us assume that the solution of eq.(S7) is obtained as $x = X(F)$. This means that the equilibrium static solution exists. Starting from this solution, we replace $F$ by the oscillating ac field $F(t) = F\cos\omega t$ to obtain

$$\dot{x}(t) = \dot{F}(t)\frac{dX}{dF}|_{F=F(t)} = -F\omega\sin\omega t\frac{dX}{dF}|_{F=F(t)}. \qquad (S8)$$

Therefore, the $\omega$-component of the imaginary part of the impedance is

$$\propto \int_0^{2\pi/\omega} dt\,(\sin\omega t)^2\frac{dX}{dF}|_{F=F(t)}. \qquad (S9)$$

With a physically reasonable assumption that $\frac{dX}{dF} > 0$, one can conclude that the inductance is negative in the limit $\omega \to 0$. However, it is not the case once the static solution $x = X(F)$ cannot be found. This corresponds to the case where the strength of $F$, the current density in the present context, is beyond the critical value to make the collective mode depinned. At finite $\omega$, the inductance can be positive as $F$ is increased even when the collective mode remains pinned due to the reduced effective $\omega_0$.

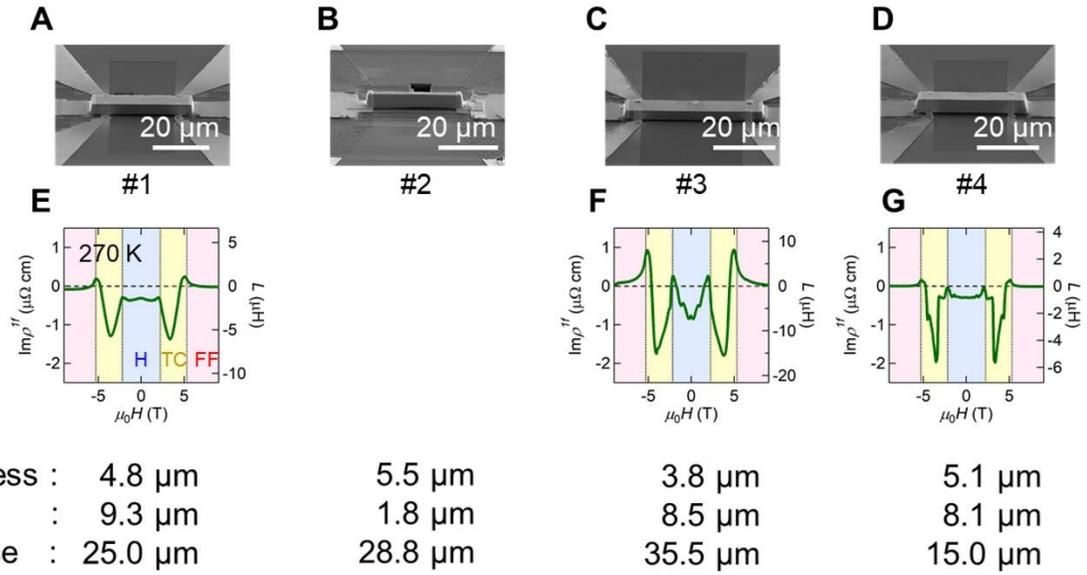

**Fig. S1.**
**Various devices for emergent inductance measurements.**
**A-D**, Scanning electron microscope images for devices with different sizes and electrode configurations. **E-G**, Magnetic-field dependence of the imaginary part of the ac resistivity Im$\rho^{1f}$ in each device at 270 K. The results of the device #1 and #3 are presented in Figs. 1 and 2 and Fig. 4 in the main text, respectively. The device #2 is the two-terminal device for the LCR measurement whose data are presented in Fig. 3 in the main text.

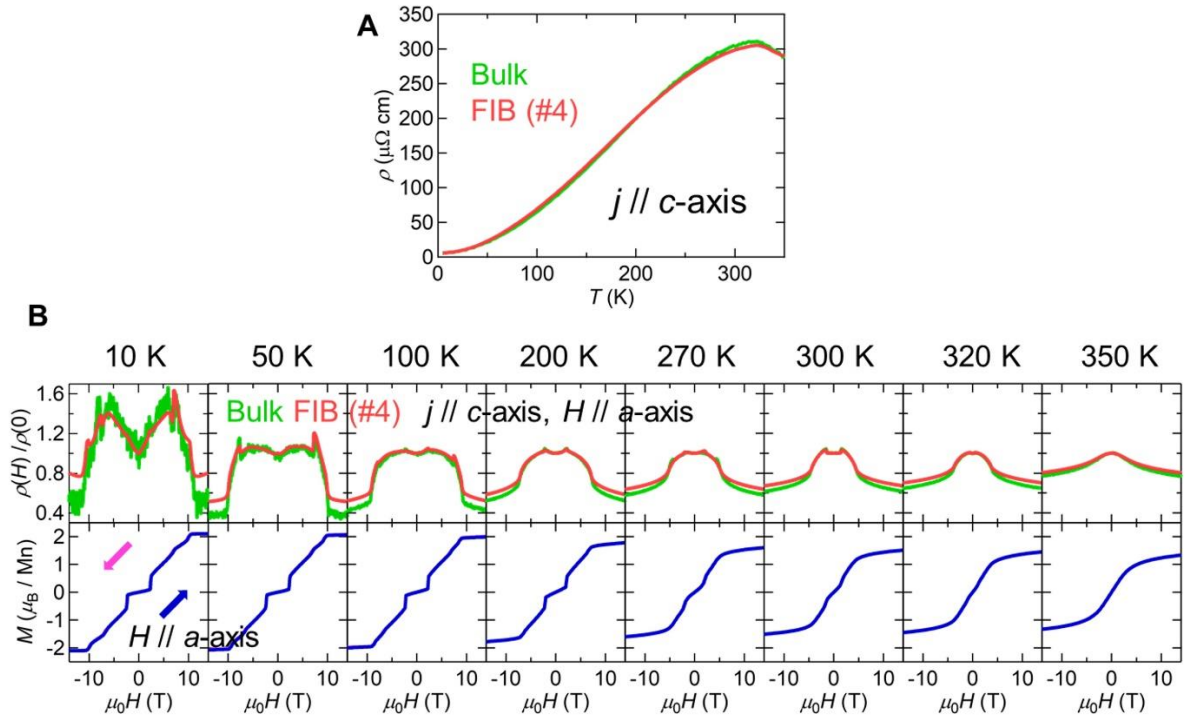

**Fig. S2.**
**Comparison of transport properties between bulk crystal and thin-plate device.**
**A**, Temperature (*T*) dependence of resistivity ($\rho$) of the bulk single crystal (green) and the thin-plate device #4 (red) of YMn$_6$Sn$_6$. **B**, Magnetic-field (*H*) dependence of resistivity $\rho(H)/\rho(0)$ (upper panels) and magnetization *M* (lower panels). No hysteresis behaviour is discerned in the magnetization curves.